\documentclass[aps,prl,superscriptaddress,showpacs,twocolumn]{revtex4-1}
\usepackage{amssymb}
\usepackage{graphicx}
\usepackage{dcolumn}
\usepackage{bm}
\usepackage{amsmath}

\usepackage[usenames]{color}
\usepackage{textcomp}
\usepackage{float}
\def\be{\begin{equation}}
\def\ee{\end{equation}}
\def\bea{\begin{eqnarray}}
\def\eea{\end{eqnarray}}
\def\nn{{\nonumber}}

\usepackage[unicode=true,bookmarks=true,bookmarksnumbered=false,bookmarksopen=false,breaklinks=false,pdfborder={0 0 1},backref=false,colorlinks=true]{hyperref}

\hypersetup{linkcolor=magenta,urlcolor=blue,citecolor=blue,pdfstartview={FitH},hyperfootnotes=false,unicode=true}

\begin{document}
\title{Spin-Induced Orbital Frustration in a Hexagonal Optical Lattice}
\author{Yongqiang Li}
\email{li\_yq@nudt.edu.cn}
\affiliation{Department of Physics, National University of Defense Technology, Changsha 410073, P. R. China}
\affiliation{Department of Physics, Graduate School of China Academy of Engineering Physics, Beijing 100193, P. R. China}
\author{Jianmin Yuan}
\affiliation{Department of Physics, National University of Defense Technology, Changsha 410073, P. R. China}
\affiliation{Department of Physics, Graduate School of China Academy of Engineering Physics, Beijing 100193, P. R. China}
\author{Xiaoji Zhou}
\affiliation{State Key Laboratory of Advanced Optical Communication System and Network, Department of Electronics, Peking University, Beijing 100871, China}
\author{Xiaopeng Li}
\email{xiaopeng\_li@fudan.edu.cn}
\affiliation{State Key Laboratory of Surface Physics, Institute of Nanoelectronics and Quantum Computing, and Department of Physics, Fudan University, Shanghai 200438, China}
\affiliation{Shanghai Qi Zhi Institute, AI Tower, Xuhui District, Shanghai 200232, China}

\begin{abstract}

Complex lattices provide a versatile ground for fascinating quantum many-body physics.
Here, we propose an exotic mechanics for generating orbital frustration in hexagonal lattices. We study two-component (pseudospin-$1/2$) Bose gases in $p$-orbital bands of two-dimensional hexagonal lattices, and find that the system exhibits previously untouched orbital frustration as a result of the interplay of spin and orbital degrees of freedom, in contrast to normal Ising-type orbital ordering of spinless $p$-orbital band bosons in two-dimensional hexagonal lattices.
Based on the classification by symmetry analysis, we find the interplay of orbital frustration and strong interaction leads to exotic Mott and superfluid phases with spin-orbital intertwined orders, in spite of the complete absence of spin-orbital interaction in the Hamiltonian.
Our study implies many-body correlations in a multi-orbital setting could induce rich spin-orbital intertwined physics in complex lattice structures.

\end{abstract}

\date{\today}


\maketitle

{\it Introduction}.
Frustration, caused by lattice geometry or long-range interactions~\cite{semeghini2021probing}, often demonstrates new emergent structures for strongly correlated quantum matter, where exotic phases appear, such as spin liquid phases~\cite{2010Balents, RevModPhys.85.1473, PhysRevLett.120.187201}, and topological states~\cite{2012Universal}. In these complex magnetic systems, a spin cannot find an orientation which simultaneously favors all the spin-spin interactions with its neighbors, as a result of the spin frustration of anti-ferromagnetic spin states~\cite{1977Toulouse}. In addition to spin, other fundamental physical quantities, such as orbital degree of freedom, can show frustration in a complex lattice as well, which provides an opportunity to investigate new orbital physics~\cite{struck2011quantum,PhysRevA.74.013607,PhysRevLett.97.190406,PhysRevB.87.224505, PhysRevLett.97.110405,2016_Li_Liu_RPP,lewenstein2011optical,PhysRevLett.121.015303}. More interestingly, the combining of spin and orbital degrees of freedom in many-body systems~\cite{book_Chaikin_1995,RevModPhys.87.457, 2007_Lewenstein_AP,2008_Bloch_Dalibard_RMP,2015_Lewenstein_RPP,RevModPhys.83.1523,Goldman_2014,RevModPhys.91.015005} potentially supports unconventional frustrations and exotic phenomena in complex lattices.

\begin{figure}[h!]
\includegraphics[trim = 0mm 0mm 0mm 0mm, clip=true, width=0.45\textwidth]{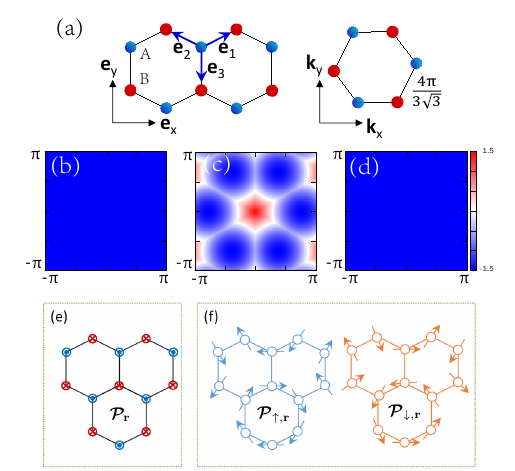}
\caption{(Color online) (a) The geometry of the two-dimensional hexagonal lattice with lattice vector ${\bf e}_m$ (left), and the first Brillouin zone (right). (b-d) The lowest $p$-orbital band of the noninteracting spinful bosonic system in a two-dimensional hexagonal lattices, with (b) $t_{\perp}=0$, (c) $t_{\parallel}=t_{\perp}$, and (d) $t_{\parallel}=0$. (e-f) Cartoons of real-space orbital polarization $\boldsymbol{\mathcal{P}}_{\sigma, {\bf r}}$ for strongly interacting many-body phases in $p$-orbital bands of the two-dimensional hexagonal lattice, where (e) spinless bosons demonstrate out-of-plane Ising-type orbital order, and (f) spinful cases inplane orbital textures, indicating spin-induced orbital frustrations.} 
\label{fig_combine}
\end{figure}
$p$-orbital ultracold atomic gases provide a new platform for investigating the interplay of spin and orbital degrees of freedom, such as spin-orbital coupling in a square lattice~\cite{PhysRevLett.121.093401,Chiral2020}, where the key element is onsite interactions for building many-body correlations. However it remains unknown how the exotic mechanism of interaction driven spin-orbit coupling carries over to other lattices, such as hexagonal lattices, where one has successfully loaded ultracold $^{87}$Rb atoms into $sp$-orbital bands of a hexagonal lattice and observed Potts-nematic superfluid~\cite{p-band_honecomb}. Generally, the orbital degree of freedom develops Ising-type order for ultracold spinless bosons in $p$-orbital band of hexagonal lattices~\cite{2016_Li_Liu_RPP}, whereas the bosons demonstrate ferromagnetic order for spin degree of freedom in many-body ground states~\cite{PhysRevLett.91.090402,altman2003phase}, indicating no frustrations for spin or orbital degree of freedom in hexagonal lattices. 

We propose a novel mechanics for generating orbital frustrations in spinful bosonic systems in hexagonal lattices. Motivated by recent achievements of $p$-orbital physics of hexagonal lattices~\cite{p-band_honecomb}, we study two-component bosonic gases loaded into the $p$-orbital bands of a two-dimensional (2D) hexagonal lattice. 
Induced by the interplay of spin and orbital degrees of freedom, we observe orbital frustrations in 2D hexagonal lattices, namely inplane circulating orbital textures, in contrast to out-of-plane Ising-type orbital ordering for the spinless case. The emergent frustrations induce exotic spin-orbital intertwined orders in both strongly interacting Mott-insulating and superfluid states, which spontaneously break time-reversal symmetry, in contrast to the time-reversal-even order established for the square lattice~\cite{PhysRevLett.121.093401}. This fascinating order can be probed with time-of-flight measurements in optical-lattice experiments.

{\it Model and Hamiltonian}.
We consider bosonic atoms prepared in two hyperfine ground states which are loaded  into the $p$-orbital bands of a 2D hexagonal optical lattice.
The corresponding annihilation operators for the bosonic particles are denoted as $p_{\nu \sigma, {\bf r}}$, with ${\bf r}$ the position of lattice sites, $\nu = x,y$ labeling the $p_x$ and $p_y$ orbital degrees of freedom, and $\sigma = \uparrow, \downarrow$ the two
pseudospin (hyperfine) states. For a compact notation, we introduce two-component spinors $\Phi_{\nu,{\bf r}}  = [p_{\nu \uparrow,{\bf r}}, p_{\nu \downarrow,{\bf r}}] ^T$. Under the single-mode approximation, the 2D hexagonal lattice can be described by a  generalized Bose-Hubbard model,
\begin{eqnarray}
\label{eq:Ham}
H &=&\sum_{m, {\bf r} \in A }  t_{\parallel}  \Phi_{m, {\bf r}} ^\dag  \Phi_{m,{\bf r} + {\bf e}_m}-\sum_{m, {\bf r} \in A} t_{\perp} \Phi^{^\prime\dagger}_{m, {\bf r}}  \Phi^\prime_{m,{\bf r} + {\bf e}_m}
+ {\rm H.c.}  \nonumber \\
&+& \frac{U_0}{2} \sum_{\bf r}   \left[ \frac{2}{3} : n_{\bf r} ^2 : -\frac{1}{3} : {\rm L}_{z,{\bf r}} ^2 :  + \frac{1}{3} :{\bf S}_{\bf r} ^2: \right]  \nonumber \\
&+&  \frac{U_2}{2}  \sum_{\bf r}\left[\frac{1}{3} :n_{\bf r}^2: -\frac{1}{3} : {\rm LS}_{z,{\bf r}} ^2: + :{\rm S}_{z,{\bf r}} ^2: - \frac{1}{3} :{\bf S}_{\bf r} ^2:  \right]\ \nonumber\\
&-&\mu \sum_{\bf r} n_{\bf r}.
\end{eqnarray}
Here, the hopping amplitudes between two nearest-neighboring $p$-orbitals along the parallel and the perpendicular directions are denoted as $t_{\parallel}$ and $t_{\perp}$, respectively.
The unit vectors ${\bf e}_{1,2}=\pm \frac{\sqrt3}{2}{\bf e}_x + \frac{1}{2}{\bf e}_y$ and ${\bf e}_3=-{\bf e}_y$ are illustrated in Fig.~\ref{fig_combine}(a).
The lattice vectors are introduced as ${\bf a}_1 = {\bf e}_1 - {\bf e}_2$, and ${\bf a}_2 = {\bf e}_1-{\bf e}_3$.
The position of $A$ and $B$ sublattices locates at ${\bf r} = l_1{\bf a}_1 + l_2{\bf a}_2 + {\bf e}_2$, and ${\bf r} = l_1{\bf a}_1 + l_2{\bf a}_2 - {\bf e}_1$, respectively, with $l_1$ and $l_2$ being integer numbers.
The lattice annihilation operators $\Phi_{m,{\bf r}} \equiv \big[(p_{{x\uparrow},{\bf r}}{\bf e}_x  + p_{{y\uparrow},{\bf r}}{\bf e}_y)\cdot {\bf e}_m ,\,\, (p_{{x\downarrow},{\bf r}}{\bf e}_x  + p_{{y\downarrow},{\bf r}}{\bf e}_y)\cdot {\bf e}_m \big]^T$ for hopping $t_{\parallel}$, and $\Phi^\prime_{m,{\bf r}} \equiv \big[( p_{{x\uparrow},{\bf r}}{\bf e}_x  + p_{y\uparrow,{\bf r}}{\bf e}_y\big)\cdot {\bf e}^\prime_m,\,\,  ( p_{{x\downarrow},{\bf r}}{\bf e}_x  + p_{y\downarrow,{\bf r}}{\bf e}_y\big)\cdot {\bf e}^\prime_m \big]^T$ with ${\bf e}^\prime_{1,2}=-\frac{1}{2}{\bf e}_x \pm \frac{\sqrt3}{2} {\bf e}_y$ and ${\bf e}^\prime_3={\bf e}_x$ for hopping $t_{\perp}$. The chemical potential $\mu$ controls the number density, and $U_{0,2}$ denotes the interaction strengths~\cite{PhysRevLett.121.093401}.
We introduce the occupation number operator,
$n_{\bf r} = \sum_{\nu} \Phi^\dag_{{\nu},{\bf r}} \Phi_{{\nu},{\bf r}}$,
the angular momentum,
${\rm L}_{z, {\bf r}}=i[\Phi_{x,{\bf r}} ^\dag \Phi_{y,{\bf r}} - \Phi_{y,{\bf r}} ^\dag \Phi_{x,{\bf r}} ]$,
the spin moment,
${\bf S}_{\bf r} = \sum_\nu \Phi^\dag_{\nu,{\bf r}} \vec{ \sigma} \Phi_{\nu,{\bf r}} $,
and a spin angular-momentum coupled operator,
${\rm LS}_{z, {\bf r}} = i[\Phi_{x,{\bf r}} ^\dag \sigma_{z} \Phi_{y,{\bf r}} - \Phi_{y,{\bf r}} ^\dag \sigma_z \Phi_{x,{\bf r}} ]$.
For convenience, we also introduce $U_{\uparrow} = U_\downarrow = U_0 + U_2$, and $U_{\uparrow\downarrow} = U_0 - U_2$, which correspond to intra- and inter-spin interactions, respectively, and are controllable in experiments via Feshbach resonances~\cite{RevModPhys.82.1225,RevModPhys.78.1311}.

We first discuss the single-particle spectrum of the noninteracting $p$-band bosonic system, described by Eq.~(\ref{eq:Ham}) with $U_{0}=U_{2}=0$. For hexagonal lattices, each unit cell consists of two nonequivalent sites, and each of which contains two $p$-orbitals, producing four bands in the system for each pseudospin $\sigma$~\cite{PhysRevLett.99.070401,PhysRevB.77.235107}. In Fig.~\ref{fig_combine}(b-d), we plot the dispersion of the lowest $p$-orbital band of the noninteracting spinful bosonic system in a 2D hexagonal lattice, with (b) $t_{\perp}=0$, (c) $t_{\parallel}=t_{\perp}$, and (d) $t_{\parallel}=0$, respectively. In the limiting case with $t_{\perp}=0$ or $t_{\parallel}=0$, the system supports a flat band over the entire Brillouin zone in the hexagonal optical lattice. The band flatness
dramatically enhances interaction effects, which has been shown to support various exotic phases~\cite{PhysRevLett.101.186807, PhysRevA.82.053618, PhysRevA.82.053611}.
In the regime of $t_{\parallel} \approx t_{\perp}$, however, the lowest $p$-orbital band is no longer rigorously flat but develops a finite width [Fig.~\ref{fig_combine}(c)].
In this dispersive band, bosons tend to form condensation. For the multi-orbital nature of this system, it is expected to develop rich spin and orbital orders in the strongly interacting limit with interaction strength comparable to the band width, which has been achieved in experiments~\cite{p-band_honecomb}.


{\it Effective orbital-exchange model.}
We focus on the physics in the strongly interacting regime with $t_{\parallel,\perp} \ll U_{\uparrow,\downarrow}$ and $U_{\uparrow\downarrow}$, and study the influence of virtual hopping processes on the orbital ordering. To quantify the orbital order, we introduce an orbital polarization vector
$\boldsymbol{\mathcal{P}}_{\sigma,{\bf r}} = \big[\langle \mathcal{{\hat P}}^x_{\sigma,\bf r}  \rangle,
\langle \mathcal{{\hat P}}^y_{\sigma,\bf r} \rangle,  \langle\mathcal{{\hat P}}^z_{\sigma,\bf r} \rangle \big]$, with $ \mathcal{{\hat P}}^x_{\sigma,\bf r} \equiv \frac{1}{2} (p_{x\sigma,{\bf r}} ^\dag p_{x \sigma, {\bf r}} - p_{y\sigma,{\bf r}} ^\dag p_{y \sigma, {\bf r}})$,
$\mathcal{{\hat P}}^y_{\sigma,\bf r} \equiv  \frac{1}{2}(p_{x\sigma,{\bf r}} ^\dag p_{y \sigma, {\bf r}} + p_{y\sigma,{\bf r}} ^\dag p_{x \sigma, {\bf r}})$, and $ \mathcal{{\hat P}}^z_{\sigma,\bf r} \equiv \frac{1}{2i}(p_{x\sigma,{\bf r}} ^\dag p_{y \sigma, {\bf r}} - p_{y\sigma,{\bf r}} ^\dag p_{x \sigma, {\bf r}})$.

For comparison, we first discuss the deep Mott physics of spinless $p$-oribtal band bosons in optical lattices with filling $n=1$, which is given by an effective orbital-exchange model
\begin{eqnarray}
H_{\rm eff} = \sum_{\langle ij \rangle }J_x \mathcal{\hat{P}}^x_i \mathcal{\hat{P}}^x_j + J_y \mathcal{\hat{P}}^y_i \mathcal{\hat{P}}^y_j + J_z \mathcal{\hat{P}}^z_i \mathcal{\hat{P}}^z_j,
\end{eqnarray}
where $\langle ij\rangle$ denotes the nearest-neighbor sites $i$ and $j$. $J_x=-3(t^2_\parallel+t^2_\perp)/2U$, $J_y=3t_\parallel t_\perp/U$, and $J_z=9t_\parallel t_\perp/U$ with $U$ being the onsite interactions between spinless bosons~\cite{2016_Li_Liu_RPP}. In the regime of $t_\parallel \approx t_\perp$, the $J_z$ term dominates, and the Ising exchange favors antiparallel configuration of nearby orbitals. This out-of-plane Ising-type order $\mathcal{P}^z_i$ survives in the hexagonal lattice~\cite{SM}. 

For the spinful bosonic gases, we focus on the spin-mixed regime with $U_{\uparrow,\downarrow} \geq U_{\uparrow\downarrow} >0$, and the lowest Mott lobe with filling $n_\uparrow+n_\downarrow=2$. In the zero-hopping limit, the single-site ground state $|\psi\rangle=\frac{1}{\sqrt2} \big(| p_{x\uparrow}, p_{y\downarrow}\rangle - | p_{x\downarrow}, p_{y\uparrow}\rangle\big)$ with $\boldsymbol{\mathcal{P}}_{\uparrow,\downarrow}=0$, indicating the absence of orbital polarization in the deep Mott-insulating regime. Away from the atomic limit but still in the strongly interacting regime, one can apply second-order perturbation theory and generate an effective Hamiltonian. To emphasis the interaction effects, we relax the superexchange processes for the spin-$\uparrow$ component by freezing the spin-$\downarrow$ one. In the regime of $t_\parallel \approx t_\perp$, the effective orbital-exchange model reads \begin{eqnarray}
H_{\rm eff} = \sum_{\langle ij \rangle }J_x \mathcal{\hat{P}}^x_{\uparrow, i} \mathcal{\hat{P}}^x_{\uparrow, j},
\end{eqnarray}
where $J_x=-\frac{\left( t_{\parallel}^{2}+t_{\bot}^{2} \right)}{K}\left[ \frac{8}{9}U_{\uparrow }^{2}+\frac{16}{3}U_{\uparrow}U_{\uparrow \downarrow}+\frac{76}{9}U_{\uparrow \downarrow}^{2} \right]$, with $K=\frac{16}{27}U_{\uparrow}^{3}+\frac{80}{27}U_{\uparrow}^{2}U_{\uparrow \downarrow}+\frac{112}{27}U_{\uparrow }U_{\uparrow \downarrow}^{2}+\frac{32}{27}U_{\uparrow \downarrow}^{3}$~\cite{SM}. For the hexagonal lattice, this effective model is transferred to $H_{\rm eff}=\sum_{m,{\bf r} \in A} J_x \mathcal{\hat{P}}_{\uparrow, {\bf r}} \mathcal{\hat{P}}_{\uparrow, {\bf r + e}_m} $ with $\mathcal{\hat{P}}_{\uparrow, {\bf r}}$ denoting the operator along the bond ${\bf e}_m$~\cite{PhysRevLett.100.160403,PhysRevLett.100.200406}, indicating the inplane orbital frustration induced by the spin degree of freedom. In other words, the system may support a phase transition from an unordered Mott insulator to an orbital ordering phase determined by frustrated orbital exchange, which is totally distinct from the spinless case in the hexagonal lattice.

{\it Frustration induced symmetry-breaking orders.}
\begin{figure}
\includegraphics[trim = 0mm 0mm 0mm 0mm, clip=true, width=0.475\textwidth]{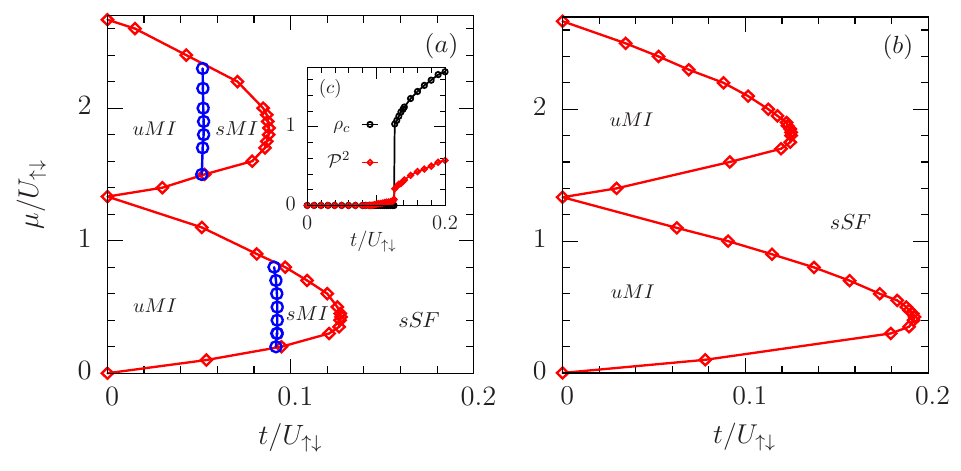}
\caption{(Color online) Filling-dependent phase diagrams of spinful bosonic gases in $p$-orbital bands of a 2D hexagonal lattice with $U_\uparrow=U_\downarrow=U_{\uparrow\downarrow}$, obtained via bosonic dynamical mean-field theory~\cite{lattice_size}. There are three quantum many-body phases with different spin-orbital intertwined orders, including the spin-orbital intertwined Mott-insulating phase (sMI) with ${\mathcal{P}}^2 \neq 0$ by breaking time-reversal symmetry, the spin-orbital intertwined superfluid phase (sSF) with both ${\mathcal{P}}^2  \neq0$ and ${\rho_c} \neq 0$ (definition in the main text) by breaking both time-reversal, $U_c(1)$ gauge, and $U_s(1)$ spin-rotational symmetries, and the unordered Mott-insulating phase (uMI) in the absence of symmetry breaking. (c) The uMI-sMI-sSF phase transitions along $\mu/U_{\uparrow\downarrow}=0.45$~\cite{lattice_size}. (a) $t \equiv t_{\parallel} = t_{\perp} $, and (b) $t \equiv t_{\parallel} = 5t_{\perp}$.}
\label{fig_mu}
\end{figure}
In order to determine the quantum ground-state orders in the spinful $p$-orbital lattice (Eq.~\eqref{eq:Ham}),  we perform a bosonic dynamical mean-field theory (BDMFT) calculation recently adapted to honeycomb lattices~\cite{PhysRevB.91.094502,PhysRevLett.120.157201} and $p$-orbital bands~\cite{PhysRevLett.121.093401}. To accommodate stripe-like orders that spontaneously break lattice-translational symmetry, we apply the real-space BDMFT~\cite{ Li2011} to our system of spinful $p$-orbital bosons in the hexagonal lattice. The technical details are described in supplementary materials~\cite{SM}.

\begin{figure}
\includegraphics[trim = 0mm 0mm 0mm 0mm, clip=true, width=0.485\textwidth]{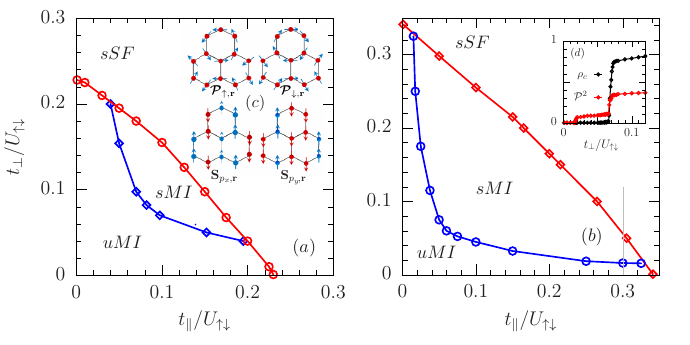}
\caption{(Color online) Hopping-dependent phase diagrams of spinful bosonic gases in $p$-orbital bands of a 2D hexagonal lattice for fixed filling $n=2$, obtained via bosonic dynamical mean-field theory~\cite{lattice_size}. The system supports the spin-orbital intertwined Mott-insulating phase (sMI), the spin-orbital intertwined superfluid phase (sSF), and the unordered Mott-insulating phase (uMI). (c) Cartoon picture of pseudospin and orbital textures of the sSF and sMI phases in real space, where emergent spin-orbital intertwined order supports circulating textures of orbital polarization $\boldsymbol{\mathcal{P}}_{\sigma,{\bf r}}$, and colinear order of pseudospin magnetism ${\bf S}_{p_\nu,{\bf r}}$. (d) The uMI-sMI-sSF phase transitions along $t_\parallel/U_{\uparrow\downarrow}=0.3$, indicated by the grey vertical line. (a) $U_\uparrow=U_\downarrow=U_{\uparrow\downarrow}$, and (b) $U_\uparrow=U_\downarrow=2U_{\uparrow\downarrow}$.}
\label{fig_txy}
\end{figure}

As expected for strongly interacting bosons in a lattice, we find both Mott insulating and superfluid phases in our system. Unlike the $s$-orbital spinful Bose-Hubbard model~\cite{PhysRevLett.91.090402}, the Mott phase of the two-component $p$-orbital bosons exhibits exotic orbital orders in the intermediate hopping regime. The orbital polarization vector $\boldsymbol{\mathcal{P}}_{\sigma,{\bf r}}$ has spatial modulations with characteristic momenta ${\bf Q}_\pm  = \pm \frac{4\pi}{9} ({\bf e}_1-{\bf e}_2)$ ($\mathcal {P}^z_{\sigma,{\bf r}}=0$), forming a inplane circulating texture structure in the hexagonal lattice \big(Fig.~\ref{fig_combine}(f) and~\ref{fig_txy}(c)\big), which is totally distinct from the spinless case~\cite{SM} \big(Fig.~\ref{fig_combine}(e)\big), indicating orbital frustration induced by spin-dependent coupling.
On the $A$- and $B$-sublattices, we have $\boldsymbol{\cal P}_{\sigma,{\bf r}} \sim s_\sigma [\cos \theta_{A, {\bf r}}, \sin \theta_{A, {\bf r}}]$ and
$\sim s_\sigma [\cos \theta_{B, {\bf r}}, \sin \theta_{B, {\bf r}}]$, respectively, with $s_{\uparrow, \downarrow}  = \pm$, and
 $\theta_{A, {\bf r}} = {\bf Q}_+ \cdot {\bf r}$,  $\theta_{B, {\bf r}} = {\bf Q}_- \cdot {\bf r}+\pi$, or the other way around. The sublattice symmetry is spontaneously broken. The orbital frustration also dominates in the strongly correlated superfluid phase, as shown in Fig.~\ref{fig_mu} and~\ref{fig_txy}.

Orbital frustration stabilizes novel long-range order in our spinful $p$-orbital lattice, where there are 16 potential spin-orbital intertwined orders, and can be classified according to their symmetry transformations~\cite{SM}. Three different types of spin-orbit intertwined orders could potentially emerge for the hexagonal lattice, including $A_1$-Odd, $A_2$-Even, and $E$-Odd, which are labeled according to their transformations under the $C_{3v}$ and ${\cal T}$ symmetries~\cite{SM}. In the Mott regime, we find the $E$-Odd order, with local observables $d^z _{x^2-y^2, {\bf r} }  \equiv \mathcal{P}^x_{\uparrow,{\bf r}} - \mathcal{P}^x_{\downarrow,{\bf r}} =  \frac{1}{2}\langle \Phi_{x,{\bf r}} ^\dag \sigma_z \Phi_{x,{\bf r}} -\Phi_{y,{\bf r}}^\dag \sigma_z \Phi_{y,{\bf r}}\rangle \neq 0 $, and $d^z _{xy, {\bf r}}  \equiv \mathcal{P}^y_{\uparrow,{\bf r}} - \mathcal{P}^y_{\downarrow,{\bf r}} = \frac{1}{2}\langle \Phi_{x,{\bf r}} ^\dag \sigma_z \Phi_{y,{\bf r}} + \Phi_{y,{\bf r}}^\dag \sigma_z \Phi_{x,{\bf r}}\rangle \neq 0$, where the phase is referred to as the spin-orbital intertwined Mott-insulator (sMI) below. This order forms a two dimensional representation ($E$) of the lattice rotation $C_{3v}$ symmetry group, and breaks the time-reversal symmetry, in sharp contrast to spin angular-momentum coupled order as established for spinful $p$-orbital bosons in the square lattice~\cite{PhysRevLett.121.093401}.
To accommodate the real-space distribution, a new order parameter is defined as $\boldsymbol{\cal P} = \boldsymbol{\cal P}_{A} + \boldsymbol{\cal P}_{B}$,
with
$\boldsymbol{\cal P}_{A} =\frac{2}{N_{\rm lat}} \sum_{{\bf r}\in A}	\sum_m {\cal O} ({\bf Q}_+ \cdot {\bf r} - {\bf Q}_+ \cdot {\bf e}_m ) [d_{x^2-y^2, {\bf r}} ^z , d_{xy, {\bf r}}^z ] ^T $ and $\boldsymbol{\cal P}_{B} =\frac{2}{N_{\rm lat}} \sum_{{\bf r}\in B}	\sum_m {\cal O} ({\bf Q}_- \cdot {\bf r} + {\bf Q}_- \cdot {\bf e}_m ) [d_{x^2-y^2, {\bf r}} ^z , d_{xy, {\bf r}} ^z ] ^T $,
where ${\cal O} (\ldots)$ represents a counter-clockwise rotation matrix by an angle $(\ldots)$, and
$N_{\rm lat}$ is the total number of lattice sites.
Its square, ${\cal P}^2$, is a $C_{6}$ rotation invariant, which characterizes the strength of the spin-orbital intertwined order, as shown in Fig.~\ref{fig_mu}(c) and \ref{fig_txy}(d).

\begin{figure}
\includegraphics[trim = 0mm 0mm 0mm 0mm, clip=true, width=0.485\textwidth]{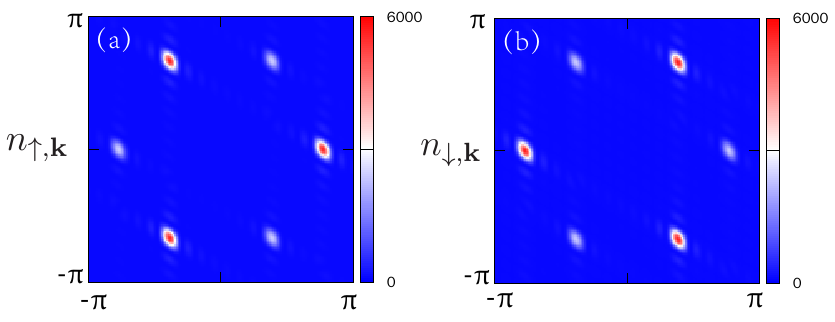}
\caption{(Color online) Momentum-space distributions of density $n_{\sigma,{\bf k}}$ for (a) pseudospin $\sigma=\uparrow$ and (b) $\downarrow$, respectively, in the strongly correlated sSF phase of spinful bosonic gases in $p$-orbital bands of a 2D hexagonal lattice, obtained via bosonic dynamical mean-field theory. The parameters are $t_{\parallel}=0.18$, $ t_{\perp}=0.1$, and $U_\uparrow$=$U_\downarrow$=$U_{\uparrow\downarrow}$=1.
}
\label{fig_momentum}
\end{figure}

In the superfluid regime, besides the $d^z_{x^2-y^2, {\bf r}}$ and $d^z_{xy, {\bf r}}$ orders, we also
find that $d^{x(y)}_{x^2-y^2, {\bf r}}  \equiv \frac{1}{2}\langle \Phi_{x,{\bf r}} ^\dag \sigma_{x(y)} \Phi_{x,{\bf r}} -\Phi_{y,{\bf r}}^\dag \sigma_{x(y)} \Phi_{y,{\bf r}} \rangle \neq 0$
and $d^{x(y)}_{xy, {\bf r}}  \equiv \frac{1}{2}\langle \Phi_{x,{\bf r}} ^\dag \sigma_{x(y)} \Phi_{y,{\bf r}} + \Phi_{y,{\bf r}}^\dag \sigma_{x(y)} \Phi_{x,{\bf r}}\rangle\neq 0$,
as a consequence of spontaneous symmetry breaking of $U_c(1)$ and $U_s(1)$, where the many-body phase is denoted as the spin-orbital intertwined superfluid phase (sSF).
The local orders in the sSF phase are still $E$-Odd orders, which correspond to the $E$ representation of the lattice $C_{3v}$ symmetry group, and remain time-reversal odd~\cite{SM}, in distinction to the emergent ordering as found in the square lattices~\cite{PhysRevA.74.013607,2009Unconventional,2016_Li_Liu_RPP,PhysRevLett.121.093401}. In addition, we find the bosons in the sSF phase condense into a state that is a linear combination of single-particle
states at two inequivalent momenta ${\bf Q}_{+}$ and ${\bf Q}_{-}$, as shown in Fig.~\ref{fig_momentum}(a)(b), indicating lattice-translational-symmetry breaking. The corresponding superfluid order $\phi_{\nu \sigma , {\bf r}}  \equiv \langle p_{\nu\sigma,{\bf r}}\rangle$ takes a form of
\bea
\phi_{\nu \sigma , {\bf r}}  = \sqrt{\frac{\rho_c}{1+\gamma^2} }
\left\{
\begin{array}{cc}
  \lambda _{\nu} ^{s_{\sigma}}  e^{i s_\sigma {\bf Q}_+ \cdot {\bf r} } + \gamma \lambda_{\nu}^ {-s_{\sigma}}   e^{is_\sigma{\bf Q}_- \cdot{\bf r}}  & \text{${\bf r} \in A$}  \nn \\
    \lambda _{\nu} ^{-s_{\sigma}}  e^{i s_\sigma {\bf Q}_+ \cdot {\bf r} } - \gamma \lambda_{\nu}^ {s_{\sigma}}   e^{is_\sigma{\bf Q}_- \cdot{\bf r}}  & \text{${\bf r} \in B$}  \nn \\
  \end{array}.
\right.
\label{eq:superfluidorder}
\eea
Here, $\rho_c \equiv \sum_{\nu\sigma,{\bf r}}|\phi_{\nu\sigma,{\bf r}}|^2/N_{\rm lat}$ represents the averaged number of condensed particles per lattice site. We have $\lambda_x ^+ = 1/2$,  $\lambda_y^+ = i/2$,
and $\lambda_\nu^- = [\lambda_\nu^+]^*$, to accommodate the  local $p_x\pm ip_y$ character of the Bloch function at the ${\bf Q}_\pm$-momentum points, and $\gamma$ is a real positive number to be consistent with the orbital order shown in Fig.~\ref{fig_mu} and~\ref{fig_txy}.

{\it Many-body phase diagrams.}
We summarize our zero-temperature phase diagrams of spinful $p$-orbital bosons of 2D hexagonal lattices in Fig.~\ref{fig_mu}, and~\ref{fig_txy}, obtained by bosonic dynamical mean-field theory. Here, averaged particle number of the condensate $\rho_c$, and spin-orbital intertwined order ${\mathcal{P}}^2$, are utilized to identify the quantum phase transitions. Without loss of generality, we mainly focus on the case with interactions $U_\uparrow=U_\downarrow=U_{\uparrow\downarrow}$, which are good approximations for $^{87}$Rb atoms.

Our calculated filling-dependent phase diagrams are presented in Fig.~\ref{fig_mu}, as a function of chemical potential and hopping amplitudes (a) $t\equiv t_{\parallel}=t_{\perp}$, and (b) $t\equiv t_{\parallel}=5t_{\perp}$, respectively.
In the lower hopping regime, two Mott-insulating phases are found: an unordered Mott insulator (uMI) in the absence of any symmetry breaking, and a sMI phase with time-reversal $\mathcal T$ and $SU(2)$ spin-rotational symmetries being spontaneously broken (the $U_s(1)$ spin-rotational symmetry develops as a subgroup instead).
In the larger hopping but still strongly interacting regime, the atoms are delocalized and a sSF phase is energetically favorable, which breaks time-reversal $\mathcal{T}$, $U_s(1)$ spin-rotational, and $U_c(1)$ gauge symmetries.
In Fig.~\ref{fig_mu}(c), the order parameters as a function of hopping amplitudes are presented to characterize the uMI-sMI-sSF transitions for $\mu/U_{\uparrow\downarrow} = 0.45$. We remark here that the phase transitions can be continues or discontinues, depending on the parameters of the model, which is beyond the scope of the present work.

Normally, the hopping amplitudes of $p$-orbital band bosons can be tuned separated in the perpendicular and parallel directions, respectively. The hopping-dependent phase diagrams are shown in Fig.~\ref{fig_txy}(a)(b)  for different interactions with fixed filling $n = 2$. We observe that there are also three many-body quantum phases for the parameter regime studied here, including the uMI, sMI, and sSF phases, where the spin-orbital intertwined phases (sMI and sSF) are stable and explore a large part of the phase diagrams, indicating large opportunities for experimentally observing these many-body quantum phases. Interestingly, we observe that pseudospin ordering $ {\bf S}_{p_\nu, {\bf r}} \equiv \frac{1}{2}\langle p^\dagger_{\nu\sigma,{\bf r}} F_{\sigma\sigma^\prime} p_{\nu\sigma^\prime,{\bf r}}\rangle$ of the spin-orbital intertwined phase is colinear with $\mathbb{Z}_3$ order in the hexagonal lattice, which coincides with the frustration-induced orbital textures $\boldsymbol{\mathcal P}_{\sigma,{\bf r}}$ via breaking lattice-translational symmetry, as shown in Fig.~\ref{fig_txy}(c). Here, $F_{\sigma\sigma^\prime}$ is the spin matrices for pseudospin-1/2. We also study the influence of the inter-spin interaction on the phase diagrams. As shown in Fig.~\ref{fig_txy}(b), we observe the Mott-insulating phase expands for $U_\uparrow=U_\downarrow=2U_{\uparrow\downarrow}$.

{\it Experimental proposal and detection.}
For the experimental detection, we expect the superfluid phase having the spin-orbital intertwined order is accessible to the present experiments~\cite{Hemmerich11,Hemmerich2013,PhysRevLett.114.115301,PhysRevLett.121.265301,p-band_honecomb}.
The symmetry-breaking pattern in the momentum distribution (Fig.~\ref{fig_momentum}) can be readily tested. Each atomic  pseudospin component  predominantly condense at ${\bf Q}_+$ and ${\bf Q}_-$, respectively, and the symmetry between them is spontaneously broken---the symmetry breaking pattern is opposite for the two components.
The spatial dependence of the spin-orbital intertwined order in the superfluid phase is determined by the phase coherence between the two momenta, which can be probed by Raman-assisted interference~\cite{2016_Li_Liu_RPP}.

{\it Conclusion}.
We study two pseudospin-component bosons loaded into $p$-orbital bands of a two-dimensional hexagonal lattice.
We find exotic circulating orbital textures with the two-pseudospin components rotating in opposite directions, which are stabilized by orbital frustration in hexagonal lattices. Accommodating this spontaneous orbital ordering, we find a spin-orbital intertwined order in both superfluid and Mott insulating phases, which corresponds to the two-dimensional $E$ representation of the lattice $C_{3v}$ group, and is time-reversal odd. Our work opens up wide opportunities for spontaneous spin-orbital coupled physics in hexagonal optical lattices. For example, the interplay of spin degrees of freedom with the renormalization stabilized Potts nematicity in an $sp^2$-orbital hybridized hexagonal lattice~\cite{p-band_honecomb} is expected to support even more exotic spontaneous spin-orbital coupled phenomena.

\textit{Acknowledgements.}
We acknowledge helpful discussions with Zhi-Fang Xu, Erhai Zhao, Rui Cao, and Hui Tan.
This work is supported by the National Natural Science Foundation of China under Grants No. 12074431, 11304386, 11774428, and 11934002, National Program on Key Basic Research Project of China (Grant No. 2017YFA0304204 and 2016YFA0301501),  and Shanghai Municipal Science and Technology Major Project (Grant No. 2019SHZDZX01).
The work was carried out at National Supercomputer Center in Tianjin, and the calculations were performed on TianHe-1A.

\bibliography{references}

\begin{widetext}
\begin{center}
	\begin{center}
	{\Large \bf Supplementary Material: Spin-Induced Orbital Frustration in a Hexagonal Optical Lattice}
\end{center}
\end{center}
\renewcommand{\theequation}{S\arabic{equation}}
\renewcommand{\thesection}{S-\arabic{section}}
\renewcommand{\thefigure}{S\arabic{figure}}
\renewcommand{\bibnumfmt}[1]{[S#1]}
\renewcommand{\citenumfont}[1]{S#1}
\setcounter{equation}{0}
\setcounter{figure}{0}



\section{bosonic dynamical mean-field theory}
The long-range order of many-body ground-state states of bosonic gases in $p$-orbital bands loaded into a 2D hexagonal optical lattice, described by Eq.~(1) in the main text, can be obtained from diagonal Green's functions $G_{\mu\nu,ij}(\tau,\tau^\prime) = \langle p_{\mu,{\bf r}_i}(\tau) p^\dagger_{\nu,{\bf r}_j}(\tau^\prime)\rangle$, and off-diagonal Green's functions $G_{\mu\nu,ij}(\tau,\tau^\prime) = \langle p_{\mu,{\bf r}_i}(\tau) p_{\nu,{\bf r}_j}(\tau^\prime)\rangle$. For the essential four-component bosonic system studied here, they are 32 independent Green's functions. In practice, we can combine different Green's functions to obtain the long-range orders listed in Table I.

To obtain these Green's functions, we utilize a bosonic version of dynamical mean-field theory (BDMFT) on the two-dimensional situation. BDMFT has been developed to provide a non-perturbative description of zero- and finite-temperature properties of the Bose-Hubbard model~\cite{Vollhardt, Hubener, Werner,Li2011,Li2012,Li2013, Liang15, Li2016, PhysRevLett.121.093401}, whose reliability of this approach has been compared against the quantum Monte-Carlo simulations~\cite{QMC_boson}. 
Actually, BDMFT has already been implemented in the honeycomb-lattice structure to study the bosonic Haldane model~\cite{PhysRevB.91.094502} and Kane-Mele-Hubbard model~\cite{PhysRevLett.120.157201}. Recently, a four-component bosonic dynamical mean-field theory has been developed to study the multi-species bosonic system in $p$-orbital bands of a two-dimensional square lattice~\cite{PhysRevLett.121.093401}. In this paper, we modify our method to study the bosonic ultracold gases loaded into the frustrated hexagonal lattices.

As in fermionic dynamical mean-field theory~\cite{georges96}, the main idea of the BDMFT approach is to map the quantum lattice problem with many degrees of freedom onto a single site. To obtain a proper local model that is approximate to the original lattice Hamiltonian, we perform an effective integration over all off-site degrees of freedom and keep only terms of subleading order. Finally, we find that the local Hamiltonian is given by a bosonic Anderson impurity Hamiltonian
\begin{eqnarray}
\hat{H}^{A}_{\bf r} &=& - \sum_{\sigma} t_\sigma \Big(\phi^{*}_{\sigma} \hat{b}_{\sigma,{\bf r}} + {\rm H.c.} \Big) + \frac{U_0}{6}  \left[ : 2n^2_{\bf r} : - : {\rm L}_{z, \bf r}^2 :  + :{\bf S}_{\bf r} ^2: \right]
+ \frac{U_2}{6}  \left[ :n^2_{\bf r}: -: {\rm LS}_{z,\bf r}^2: + :3{\rm S}_{z,{\bf r}} ^2: - :{\bf S}_{\bf r} ^2:  \right] -\mu n_{\bf r}\nonumber
\\
&+&  \sum_{l}  \epsilon_l \hat{a}^\dagger_l\hat{a}_l + \sum_{l,\sigma} \Big( V_{\sigma,l} \hat{a}_l\hat{b}^{\dagger}_{\sigma,{\bf r}} + W_{\sigma,l} \hat{a}_l\hat{b}^{}_{\sigma,{\bf r}} + {\rm H.c.} \Big),
\end{eqnarray}
where $:\ldots:$ denotes normal ordering, $t_\sigma$ is the nearest-neighbor hopping amplitude for the four components denoted as $\sigma$, the chemical potential and interaction terms are directly inherited from the Hubbard Hamiltonian. The bath of condensed bosons is represented by the Gutzwiller term with
superfluid order parameters $\phi_{\sigma}$ for the component $\sigma$. The bath of normal bosons is
described by a finite number of orbitals with creation operators $\hat{a}^\dagger_l$ and energies
$\epsilon_l$, where these orbitals are coupled to the impurity via normal-hopping amplitudes
$V_{\sigma, l}$ and anomalous-hopping amplitudes $W_{\sigma, l}$. The anomalous hopping terms are
needed to generate the off-diagonal elements of the hybridization function. Note here that we use ${\hat b}_\sigma$ to denote the four-component bosonic annihilation operator, to shorten the notation of the function.

To solve the Anderson Hamiltonian, we apply exact diagonalization as a solver to solve the local problem~\cite{Caffarel_1994, georges96}. After diagonalization, we obtain the eigenstates, eigenenergies, and local Green's functions in the Lehmann-representation
\begin{eqnarray}
G_{A,\sigma \sigma'}^1 (i \omega_n) &=& \frac{1}{Z} \sum_{mn} \langle m | \hat b_\sigma | n\rangle \langle n | \hat b_{\sigma'}^\dagger | m \rangle \frac{e^{- \beta E_n} - e^{-\beta E_m}}{E_n - E_m + i \hbar \omega_n} + \beta \phi_\sigma \phi^\ast_{\sigma'} \\
G_{A,\sigma \sigma'}^2 (i \omega_n) &=& \frac{1}{Z} \sum_{mn} \langle m | \hat b_\sigma | n\rangle \langle n | \hat b_{\sigma'} | m \rangle \frac{e^{- \beta E_n} - e^{-\beta E_m}}{E_n - E_m + i \hbar \omega_n} + \beta \phi_\sigma \phi_{\sigma'},
\end{eqnarray}
where $\omega_n$ denotes Matsubara frequency, the superfluid order parameter $\phi_\sigma = \langle\hat{b}_\sigma\rangle_{0}$, and the notation $\langle \ldots \rangle_0$ means that the expectation value is calculated in the cavity system~\cite{Hubener}.

Then, the local self energy for each site can obtained via Dyson equation:
\begin{eqnarray}
\mathbf{\Sigma}_{A}(i\omega_n)= \boldsymbol{\mathcal{G}}^{-1}_{A}(i\omega_n)- \mathbf{G}^{-1}_{A}(i\omega_n),
\end{eqnarray}
where $\boldsymbol{\mathcal{G}}_{A,\sigma\sigma'}(i\omega_n)$ denotes the noninteracting Weiss Green's function of the Anderson impurity site.

After obtaining the self energy for all the sites, we can employ the Dyson equation in real-space representation to compute the interacting lattice Green's function:
\begin{eqnarray}
\mathbf{G}^{-1}(i\omega_n)= \mathbf{G}^{-1}_0(i\omega_n) - \mathbf{\Sigma}(i\omega_n),
\label{lattice_G}
\end{eqnarray}
where the noninteracting lattice Green's function $\mathbf{G}^{-1}_0(i\omega_n) = (\mu+ i\omega_n){\bf 1} - {\bf t} $, with the matrix of hopping $\bf t$ determined by lattice structures. Note here that the site-dependence of the Green's functions and self energies are shown by boldface quantities that denote a matrix form with site-indexed elements.

The self-consistency loop is solved as follows: starting from an initial choice for the Anderson parameters and the superfluid order parameters, the Anderson Hamiltonian is constructed in the Fock basis and diagonalized to obtain the eigenstates and eigenenergies. The eigenstates and energies allow us to calculate the superfluid order parameter, the impurity Green's functions and self-energies, and then obtain the lattice Green's functions via Eq.~(\ref{lattice_G}). Subsequently, new Anderson parameters are obtained, by comparing the new Green's functions with the old ones. With these new Anderson parameters, the procedure is iterated until convergence is reached.

\section{Symmetry analysis}
\begin{table}[tp]
    \caption{Local symmetry-breaking orders for two-component bosons loaded into $p$-orbital bands of a hexagonal lattice. The bilinear operators $\Phi_{\nu} ^\dag \sigma_z \Phi_{\nu'}$ keep the $U_s(1)$ spin-rotational symmetry preserved, whereas the operators $\Phi_{\nu} ^\dag \sigma_{x,y} \Phi_{\nu'}$ break this symmetry. Note here that we omit the lattice position $\bf r$ in the operator $\Phi_{\nu,{\bf r}}$, to shorten the donation.}
    \begin{tabular}{|c|c|c| }
    \hline \nonumber
    Operators    							& $C_{3 v}$ 	&${\cal T}$     \\\hline\hline
    $\Phi_x^\dag \Phi_x + \Phi_y^\dag \Phi_y$ 	& $A_1$ 		&Even 		\\ \hline
    $i[\Phi_x ^\dag \Phi_y -\Phi_y ^\dag \Phi_x]$ 	& $A_2$ 		&Odd 		\\ \hline
    $\left\{ \begin{array}{c}
    		 \Phi_x ^\dag \Phi_x -\Phi_y^\dag \Phi_y \\
		 \Phi_x ^\dag \Phi_y + \Phi_y^\dag \Phi_x
		 \end{array} \right. $
		 							& $E$		& Even 		\\  \hline
    $\Phi_x ^\dag \sigma_z \Phi_x + \Phi_y^\dag \sigma_z \Phi_y$ ; $\Phi_x ^\dag \sigma_{x,y} \Phi_x + \Phi_y^\dag \sigma_{x,y} \Phi_y$
    									& $A_1$ 		&Odd 		\\ \hline
    $i[\Phi_x ^\dag \sigma_z \Phi_y-\Phi_y ^\dag \sigma_z \Phi_x]$ ; $  i[\Phi_x ^\dag \sigma_{x,y} \Phi_y-\Phi_y ^\dag \sigma_{x,y} \Phi_x]$
    									&$A_2$		&Even 		\\ \hline												
      $\left\{ \begin{array}{c}
    		 \Phi_x ^\dag \sigma_z \Phi_x -\Phi_y^\dag \sigma_z \Phi_y \\
		 \Phi_x ^\dag \sigma_z \Phi_y + \Phi_y^\dag \sigma_z \Phi_x
		 \end{array} \right. $ ;
      $\left\{ \begin{array}{c}
    		 \Phi_x ^\dag \sigma_{x,y} \Phi_x -\Phi_y^\dag \sigma_{x,y} \Phi_y \\
		 \Phi_x ^\dag \sigma_{x,y} \Phi_y + \Phi_y^\dag \sigma_{x,y} \Phi_x
		 \end{array} \right. $
		 							& $E$		& Odd 		\\  \hline 		 							
   \hline
    \end{tabular}
\label{tab:order}
\end{table}
For the complexity of our spinful $p$-orbital lattice, we classify the potential long-range orders according to their symmetry transformations. The system has a $U_c (1)$ phase symmetry associated with particle-number conservation in $\sum_{\nu, {\bf r}} \Phi_{\nu,{\bf r}}^\dag   \Phi_{\nu,{\bf r}}$, and a $U_s(1)$ spin-rotational symmetry associated with spin conservation in $\sum_{\nu, {\bf r}}  \Phi_{\nu,{\bf r}}^\dag  \sigma_z \Phi_{\nu,{\bf r}}$. Apart from these continuous symmetries, it also has discrete lattice rotation  $C_{3v}$, and time-reversal ${\cal T}$ symmetries.
Different symmetry-breaking orders in this system would give rise to local quadratic observables of different symmetry properties listed in Table~\ref{tab:order}.
The time-reversal invariant spin angular-momentum intertwined order
${\rm LS}_{z,{\bf r}} = i[\Phi_{x,{\bf r}} ^\dag \sigma_z \Phi_{y,{\bf r}}-\Phi_{y,{\bf r}} ^\dag \sigma_z \Phi_{x,{\bf r}}]$ has been found for spinor bosons loaded into $p$-orbital bands of a square lattice~\cite{PhysRevLett.121.093401}.
For the hexagonal lattice, three different types of spin-orbit intertwined orders could potentially emerge, including $A_1$-Odd, $A_2$-Even, and $E$-Odd, which are labeled according to their transformations under the $C_{3v}$ and ${\cal T}$ symmetries, as shown in Table~\ref{tab:order}.

\section{Effect orbital-exchange model of spinful bosonic gases in $p$-orbital bands of a hexagonal lattice}
In this part, we derive the effective orbital-exchange model for the spinful bosonic gases in $p$-orbital bands of a hexagonal lattice. We consider the lowest Mott lobe with filling $n=n_\uparrow+n_\downarrow=2$ trapped in a single site. In this atomic limit with $t_\parallel=t_\perp=0$, the system is described by
\begin{small}
\begin{equation}
\begin{aligned}
H_0&=\frac{1}{2}\sum_{\sigma \sigma \prime,\nu,{\bf r}}{U_{\sigma \sigma \prime}b_{\nu \sigma}^{\dagger}\left( {\bf r}\right) b_{\nu \sigma \prime}^{\dagger}\left( {\bf r} \right) b_{\nu \sigma \prime}\left( {\bf r} \right) b_{\nu \sigma}\left( {\bf r} \right)}+\frac{1}{2}\sum_{\sigma \sigma \prime,\mu \ne \nu, {\bf r}}{V_{\sigma \sigma \prime}}\left[ b_{\mu \sigma}^{\dagger}\left( {\bf r} \right) b_{\mu \sigma \prime}^{\dagger}\left( {\bf r} \right) b_{\nu \sigma \prime}\left( {\bf r} \right) b_{\nu \sigma}\left( {\bf r} \right) \right. \\
&\left. +b_{\mu \sigma}^{\dagger}\left( {\bf r} \right) b_{\mu \sigma}\left( r \right) b_{\nu \sigma \prime}^{\dagger}\left( {\bf r} \right) b_{\nu \sigma \prime}\left( {\bf r} \right) +b_{\mu \sigma}^{\dagger}\left( {\bf r} \right) b_{\mu \sigma \prime}\left( {\bf r} \right) b_{\nu \sigma \prime}^{\dagger}\left( {\bf r} \right) b_{\nu \sigma}\left( {\bf r} \right) \right].
\end{aligned}
\end{equation}
\end{small}

We construct the basis as $|p_{\nu\sigma}, p_{\nu^\prime\sigma^\prime}\rangle$, where $\nu=x,y$ and $\sigma=\uparrow,\downarrow$, under which the above Hamiltonian can be expressed as
\begin{small}
\begin{equation}
\begin{aligned}
H_0=\left( \begin{matrix}	U_{\uparrow \downarrow}&		&		&		&		&		&		&		 V_{\uparrow \uparrow}&		&		\\	&		 U_{\uparrow \uparrow}&		&		&		&		&		&		 &		V_{\uparrow \downarrow}&		\\	&		&		2V_{\uparrow \uparrow}&		&		&		&		 &		&		&		\\	&		&		&		V_{\uparrow \downarrow}&		&		V_{\uparrow \downarrow}&		&		&		&		\\	&		&		&		&		U_{\downarrow \downarrow}&		 &		&		&		&		 V_{\downarrow \downarrow}\\	&		&		&		V_{\uparrow \downarrow}&		 &		V_{\uparrow \downarrow}&		&		&		&		\\	 &		&		&		&		&		&		 2V_{\downarrow \downarrow}&		&		&		\\	V_{\uparrow \uparrow}&		&		&		&		 &		 &		&		U_{\uparrow \downarrow}&		&		\\	&		V_{\uparrow \downarrow}&		&		&		 &		&		&		&		 U_{\uparrow \downarrow}&		\\	&		&		&		&		 V_{\downarrow \downarrow}&		&		&		&		&		U_{\downarrow \downarrow}\\\end{matrix} \right).
\end{aligned}
\end{equation}
\end{small}

After diagonalizable the matrix, we obtain the eigenstates
\begin{equation}
\begin{aligned}
|\psi_1\rangle&=\frac{1}{\sqrt{2}}\left( \left| p_{x\uparrow},p_{x\uparrow} \right> +\left| p_{y\uparrow},p_{y\uparrow} \right> \right)\\
|\psi _2\rangle&=\frac{1}{\sqrt{2}}\left( \left| p_{x\uparrow},p_{x\downarrow} \right> +\left| p_{y\uparrow},p_{y\downarrow} \right> \right)\\
|\psi _3\rangle&=\left| p_{x\uparrow},p_{y\uparrow} \right>\\
|\psi _4\rangle&=\frac{1}{\sqrt{2}}\left( \left| p_{x\uparrow},p_{y\downarrow} \right> +\left| p_{x\downarrow},p_{y\uparrow} \right> \right)\\
|\psi _5\rangle&=\frac{1}{\sqrt{2}}\left( \left| p_{x\downarrow},p_{x\downarrow} \right> +\left| p_{y\downarrow},p_{y\downarrow} \right> \right)\\
|\psi _6\rangle&=\left| p_{x\downarrow},p_{y\downarrow} \right>\\
|\psi _7\rangle&=\frac{1}{\sqrt{2}}\left( \left| p_{x\uparrow},p_{x\uparrow} \right> -\left| p_{y\uparrow},p_{y\uparrow} \right> \right)\\
|\psi _8\rangle&=\frac{1}{\sqrt{2}}\left( \left| p_{x\uparrow},p_{x\downarrow} \right> -\left| p_{y\uparrow},p_{y\downarrow} \right> \right)\\
|\psi _9\rangle&=\frac{1}{\sqrt{2}}\left( \left| p_{x\uparrow},p_{y\downarrow} \right> -\left| p_{x\downarrow},p_{y\uparrow} \right> \right)\\
|\psi _{10}\rangle&=\frac{1}{\sqrt{2}}\left( \left| p_{x\downarrow},p_{x\downarrow} \right> -\left| p_{y\downarrow},p_{y\downarrow} \right> \right),\\
\end{aligned}
\end{equation}
and the eigenenergies
\begin{equation}
\begin{aligned}
&\epsilon _1=\frac{1}{2}\left(U_{\uparrow \uparrow}+V_{\uparrow \uparrow}\ \right),\,   &   \epsilon _2&=\frac{1}{2}\left(U_{\uparrow \downarrow}+V_{\uparrow \downarrow}\right),\\
&\epsilon _3=V_{\uparrow \uparrow},\,\,        &     \epsilon _4&=V_{\uparrow \downarrow},\\&\epsilon _5=\frac{1}{2}\left(U_{\downarrow \downarrow}+V_{\downarrow \downarrow}\right),\,\, &     \epsilon _6&=V_{\downarrow \downarrow},\\
&\epsilon _7=\frac{1}{2}\left(U_{\uparrow \uparrow}-V_{\uparrow \uparrow}\right),\,\,   &   \epsilon _8&=\frac{1}{2}\left(U_{\uparrow \downarrow}-V_{\uparrow \downarrow}\right),\\
&\epsilon _9=0,                &    \epsilon _{10}&=\frac{1}{2}\left(U_{\downarrow \downarrow}-V_{\downarrow \downarrow}\right).\\
\end{aligned}
\end{equation}

From these eigenenergies, we find that the ground state is $|\psi _9\rangle$ in the regime of $U_{\uparrow,\downarrow} \geq U_{\uparrow\downarrow} >0$. With the above eigenvectors, we next calculate the expectation values of the operators for the ground state,
\begin{equation}
\left< \mathcal{P}_{\sigma}^{x} \right> =\left< \mathcal{P}_{\sigma}^{y} \right> =\left< \mathcal{P}_{\sigma}^{z} \right> =0,\, {\rm and}\, \left< \mathcal{P}_{\sigma}^{2} \right> =\frac{3}{4},
\end{equation}
where the orbital polarization operators are given by
\begin{equation}
\begin{aligned}
\mathcal{P}_{\sigma}^{x}&=\frac{1}{2}\left( b_{x\sigma}^{\dagger}b_{x\sigma}-b_{y\sigma}^{\dagger}b_{y\sigma}
\right)\\
\mathcal{P}_{\sigma}^{y}&=\frac{1}{2}\left( b_{x\sigma}^{\dagger}b_{y\sigma}+b_{y\sigma}^{\dagger}b_{x\sigma} \right) \\
\mathcal{P}_{\sigma}^{z}&=\frac{-i}{2}\left( b_{x\sigma}^{\dagger}b_{y\sigma}-b_{y\sigma}^{\dagger}b_{x\sigma} \right). \\
\end{aligned}
\end{equation}

To emphasis the interaction effects, we relax the superexchange processes for the spin-$\uparrow$ component by freezing the spin-$\downarrow$ one with $n_\downarrow=1$. In the low-hopping regime, one can obtain an effective orbital-exchange Hamiltonian from second-order perturbation theory
\begin{equation}
H^{\prime}_{\alpha \beta}=-\sum_{\gamma}{\frac{V_{\alpha \gamma}V_{\gamma \beta}}{E_{\gamma}-\frac{E_{\alpha}+E_{\beta}}{2}}},
\end{equation}
where $\alpha$ and $\beta$ label the ground states, $\gamma $ denotes the excited states, and $E$ are the eigenenergies of the zeroth order part $H_0$, with $V$ being the hopping terms treated as perturbation. Finally, we obtain an effective orbital-exchange Hamiltonian for the spin-$\uparrow$ component
\begin{equation}
\label{eq:effective_square}
\begin{aligned}
H_{\rm eff}&=\sum_{\langle ij \rangle} J_x \mathcal{P}_{\uparrow,i}^{x} \mathcal{P}_{\uparrow,j}^{x} + J_{xy} \left[ \mathcal{P}_{\uparrow,i}^{x} \mathcal{P}_{\uparrow,j}^{y} +\mathcal{P}_{\uparrow,i}^{y} \mathcal{P}_{\uparrow,j}^{x} \right], \\
\end{aligned}
\end{equation}
where $\langle ij\rangle$ denotes the nearest-neighbor sites $i$ and $j$ of a lattice, $J_x=-\frac{\left( t_{\parallel}^{2}+t_{\bot}^{2} \right)}{K}\left[ \frac{8}{9}U_{\uparrow \uparrow}^{2}+\frac{16}{3}U_{\uparrow \uparrow}U_{\uparrow \downarrow}+\frac{76}{9}U_{\uparrow \downarrow}^{2} \right]$, $J_{xy}=-\frac{4\left( t_{\parallel}^{2}-t_{\bot}^{2} \right) }{K} \left( \frac{2}{9}U_{\uparrow \uparrow}U_{\uparrow \downarrow}+\frac{2}{3}U_{\uparrow \downarrow}^{2} \right)$, with $K=\frac{16}{27}U_{\uparrow \uparrow}^{3}+\frac{80}{27}U_{\uparrow \uparrow}^{2}U_{\uparrow \downarrow}+\frac{112}{27}U_{\uparrow \uparrow}U_{\uparrow \downarrow}^{2}+\frac{32}{27}U_{\uparrow \downarrow}^{3}$.

To generalize to the hexagonal lattice, we can rotate $p$-orbitals at an angle of $\theta$ with the $x$ axis. Along this direction, the operators transform as $b^\prime_{x\sigma} = {\rm cos\theta}\,b_{x\sigma} + {\rm sin \theta} \, b_{y\sigma}$ and $b^\prime_{y\sigma} = -{\rm sin\theta}\, b_{x\sigma} + {\rm cos\theta}  \, b_{y\sigma}$. Accordingly, the orbital polarization operators $\mathcal{P}_{\sigma}^{x\prime} = {\rm cos 2\theta}\, \mathcal{P}_{x\sigma} + {\rm sin 2\theta} \, \mathcal{P}_{y\sigma}$, $\mathcal{P}_{\sigma}^{y\prime} = -{\rm sin 2\theta}\, \mathcal{P}_{x\sigma} + {\rm cos 2\theta} \, \mathcal{P}_{y\sigma}$, and $\mathcal{P}^\prime_z= \mathcal{P}_z$. In the regime of $t_\parallel \approx t_\perp$, the $J_x$ term dominates in the effective model~(\ref{eq:effective_square}), which is given by
\begin{equation}
\begin{aligned}
H_{\rm eff}=\sum_{{\bf e}_\theta,{\bf r} \in A} J_x \mathcal{P}^\prime_{\uparrow, {\bf r}} \mathcal{P}^\prime_{\uparrow, {\bf r + e}_\theta}
\end{aligned}
\end{equation}
for the hexagonal lattice, with $\mathcal{P}^\prime_{\uparrow, {\bf r}}$ denoting the operator along the bond ${\bf e}_\theta$ directing at angle $\theta$ with the $x$ axis.

\section{Phase diagram of spinless bosonic gases in $p$-orbital bands of a hexagonal lattice}
\begin{figure}[h]
\includegraphics[trim = 0mm 0mm 0mm 0mm, clip=true, width=0.75\textwidth]{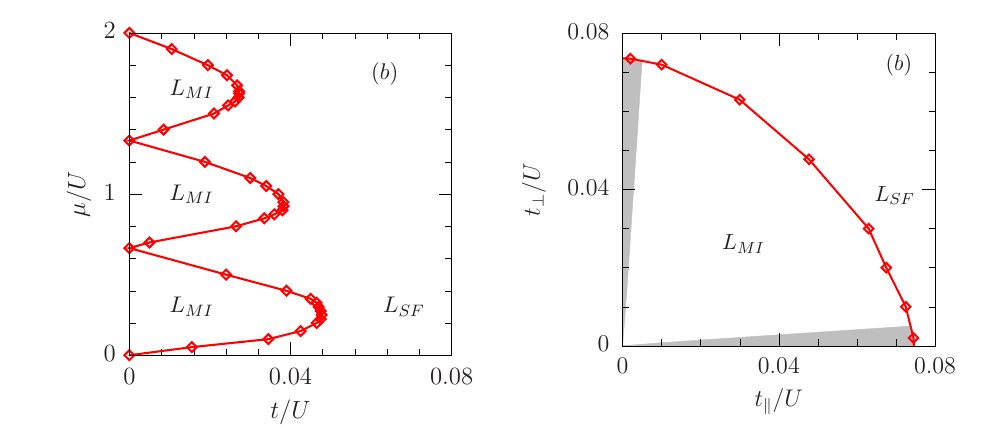}
\caption{(Color online) (a) Filling- and (b) hopping-dependent phase diagrams of spinless bosonic gases in $p$-orbital bands of a 2D hexagonal lattice, obtained via bosonic dynamical mean-field theory. Whereas the Mott-insulating ($L_{MI}$) and superfluid ($L_{SF}$) phases prefer Ising-type orbital order with $\mathcal{P}^z_{\bf r} \neq0$ in the regime $t_\parallel\approx t_\perp$, the Ising-type orbital order of the Mott phase vanishes in the regime $t_\perp \ll t_\parallel$ or $t_\parallel \ll t_\perp$, marked by the shaded region. (a) $t \equiv t_{\parallel} = t_{\perp} $, and (b) $n=1$.}
\label{fig_mu_hopping}
\end{figure}
We consider spinless bosonic atoms prepared in the hyperfine ground state which is loaded into the $p$-orbital bands of a two-dimensional (2D) hexagonal optical lattice.
The corresponding annihilation operators for the bosonic particles are denoted as $p_{\nu \sigma, {\bf r}}$, with ${\bf r}$ the position of lattice sites, $\nu = x,y$ labeling the $p_x$ and $p_y$ orbital degrees of freedom. Under the single-mode approximation, the 2D hexagonal lattice can be described by a  generalized Bose-Hubbard model,
\begin{eqnarray}
\label{eq:Ham}
H &=&\sum_{m, {\bf r} \in A }  t_{\parallel}  p_{m, {\bf r}} ^\dag  p_{m,{\bf r} + {\bf e}_m}-\sum_{m, {\bf r} \in A} t_{\perp} p^{^\prime\dagger}_{m, {\bf r}}  p^\prime_{m,{\bf r} + {\bf e}_m}
+ {\rm H.c.} + \frac{U}{2} \sum_{\bf r}   \left[ n_{\bf r} ^2  -\frac{1}{3} {\rm L}_{z,{\bf r}} ^2  \right]\ - \mu \sum_{\bf r} n_{\bf r}.
\end{eqnarray}
Here, the hopping amplitudes between two nearest-neighboring $p$-orbitals along the parallel and the perpendicular directions are denoted as $t_{\parallel}$ and $t_{\perp}$, respectively.
The unit vectors ${\bf e}_{1,2}=\pm \frac{\sqrt3}{2}{\bf e}_x + \frac{1}{2}{\bf e}_y$ and ${\bf e}_3=-{\bf e}_y$ are illustrated in Fig. 1(a) in the main text.
The lattice vectors are introduced as ${\bf a}_1 = {\bf e}_1 - {\bf e}_2$, and ${\bf a}_2 = {\bf e}_1-{\bf e}_3$.
The position of $A$ and $B$ sublattices locates at ${\bf r} = l_1{\bf a}_1 + l_2{\bf a}_2 + {\bf e}_2$, and ${\bf r} = l_1{\bf a}_1 + l_2{\bf a}_2 - {\bf e}_1$, respectively, with $l_1$ and $l_2$ being integer numbers.
The lattice annihilation operators $p_{m,{\bf r}} \equiv (p_{x,{\bf r}}{\bf e}_x  + p_{y,{\bf r}}{\bf e}_y)\cdot {\bf e}_m $ for hopping $t_{\parallel}$, and $p^\prime_{m,{\bf r}} \equiv ( p_{x,{\bf r}}{\bf e}_x  + p_{y,{\bf r}}{\bf e}_y\big)\cdot {\bf e}^\prime_m$ with ${\bf e}^\prime_{1,2}=-\frac{1}{2}{\bf e}_x \pm \frac{\sqrt3}{2} {\bf e}_y$ and ${\bf e}^\prime_3={\bf e}_x$ for hopping $t_{\perp}$. $\mu$ denotes the the chemical potential, and $U$ denotes the interaction strengths. We introduce the occupation number operator, $n_{\bf r} = \sum_{\nu} p^\dag_{{\nu},{\bf r}} p_{{\nu},{\bf r}}$, the angular momentum, ${\rm L}_{z, {\bf r}}=i[p_{x,{\bf r}} ^\dag p_{y,{\bf r}} - p_{y,{\bf r}} ^\dag p_{x,{\bf r}} ]$.

As shown in the effective spin model, the system can develop two types of long-range order in the Mott-insulating phases. We confirm this conclusion via numerical simulations, based on bosonic dynamical mean-field theory, and observe that there are three different quantum phases in the spinless bosonic gases in the $p$-orbital band of hexagonal lattices, including two Mott-insulating phases and one superfluid phase, as shown in Fig.~\ref{fig_mu_hopping}. In the low-hopping regime $t_\parallel\approx t_\perp \ll U$, the $J_z$ term dominates and the system favors a Mott phase ($L_{MI}$) with staggered orbital angular momentum $\mathcal{P}^z_{\bf r}\neq0$, and in the regime $t_\parallel \ll t_\perp$ or $t_\perp \ll t_\parallel$, the $J_x$ term dominates and a new Mott phase appears with the absence of staggered orbital angular momentum. In the larger hopping regime, the atoms delocalize and the superfluid phase ($L_{SF}$) with staggered orbital angular momentum develops.

\section{Phase diagram of heteronuclear Bose-Bose mixtures in hexagonal lattices}
\begin{figure}[h]
\includegraphics[trim = 0mm 0mm 0mm 0mm, clip=true, width=0.5\textwidth]{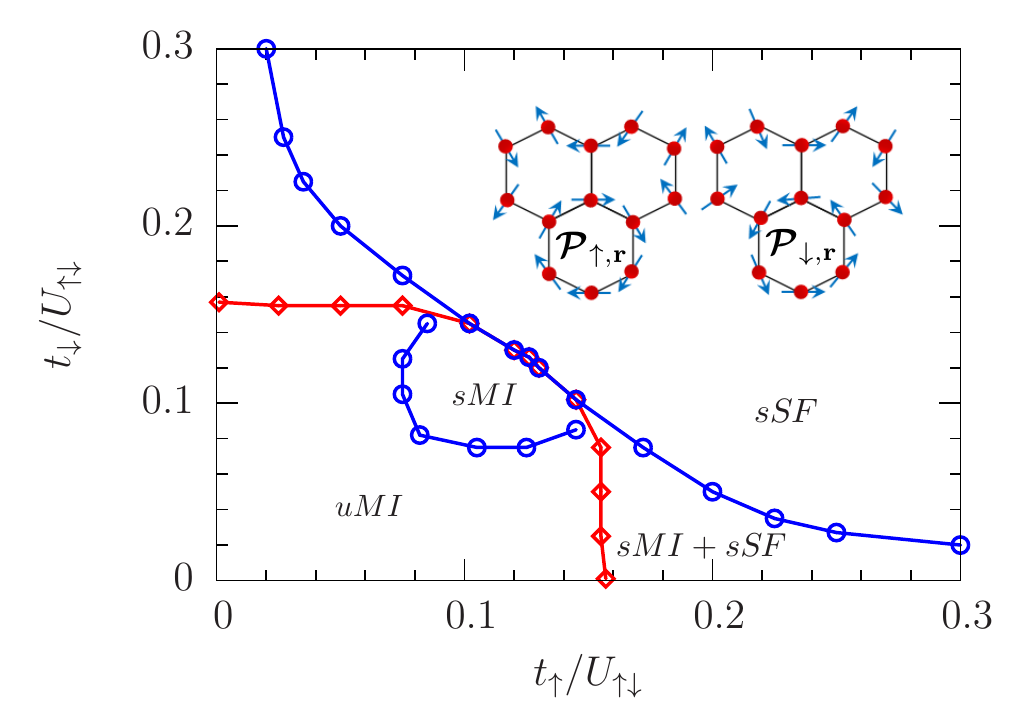}
\caption{(Color online) Spin-dependent phase diagram of bosonic mixtures in $p$-orbital bands of a 2D hexagonal lattice with $n=2$, obtained via bosonic dynamical mean-field theory. There are four phases in the system, including unordered Mott insulator (uMI), spin-orbital intertwined Mott insulator (sMI), one-component superfluid (sMI+sSF), and spin-orbital smectic superfluid (sSF). Other parameters $U_\uparrow=U_\downarrow=U_{\uparrow\downarrow}$, $t_\uparrow \equiv t_{\parallel\uparrow} = t_{\perp\uparrow}$, and $t_\downarrow \equiv t_{\parallel\downarrow} = t_{\perp\downarrow}$.}
\label{fig_RbNa}
\end{figure}
Up to now, theoretical calculations were mainly performed for the symmetric parameters with $t_\parallel$ and $t_{\perp}$ being identical for the two pseudospin components, which is the most relevant case for the present experiments. However, Bose-Bose mixtures may consist of different atomic species, such as $^{87}$Rb and $^{7}$Li, which generally do not have these symmetric properties. Without loss of generality, we still focus on $U_\uparrow=U_\downarrow=U_{\uparrow\downarrow}$ and $t_{\parallel\sigma}= t_{\perp\sigma}$, where $t_{\parallel\sigma}$ and $t_{\perp\sigma}$ are the hopping amplitudes along the parallel and perpendicular directions for pseudospin $\sigma$, respectively.

As shown in Fig.~\ref{fig_RbNa}, a phase diagram is demonstrated as a function of $t_\uparrow\equiv t_{\parallel\uparrow} = t_{\perp\uparrow}$ and $t_\downarrow =  t_{\parallel\downarrow} = t_{\perp\downarrow}$, where four phases appears, including unordered Mott phases (uMI), spin-orbital intertwined Mott phases (sMI), spin-orbital smectic superfluid phase (sSF), and a new phase (sMI+sSF) when the hopping amplitude for one component is small and the other one relatively large. In this new phase (sMI+sSF), we observe that one species is the Mott-insulating state, and the other one the superfluid phase. In addition, we observe exotic orbital textures $\boldsymbol{\mathcal P }_{\sigma,{\bf r}}$ for this sMI+sSF phase, with the Mott-insulating species rotating clockwise and the superfluid one anti-clockwise, or vice versa, as shown in the inset of Fig.~\ref{fig_RbNa}, indicating the spin-orbital intertwined order of the many-body ground state emerging. 

\end{widetext}
\bibliography{references}
\end{document}